\documentclass[11pt]{article}
\usepackage{cite}
\usepackage{amsmath,amsfonts,amssymb,calrsfs}
\usepackage[small,bf,hang]{caption}
\def\hybrid{
        \topmargin -20pt
        \oddsidemargin 0pt
        \headheight 0pt \headsep 0pt
        \textwidth 6.25in % A4 paper
        \textheight 9.5in % A4 paper
        \marginparwidth .875in
        \parskip 5pt plus 1pt \jot = 1.5ex}

\hybrid

 \csname
@addtoreset\endcsname{equation}{section}

\def\moth{\mathsurround=0pt}
\newdimen\zo \zo=0pt

\def\tick{\leaders\hrule height 0.5ex depth 0pt \hskip 0.5pt}
\def\upboxfill{$\moth \setbox\zo\hbox{\tick}%
  \hskip 3pt\hbox to 0pt{$\tick$\hss}\hrulefill \hbox to 7.5pt{$\tick$\hss}$}

\def\dtick{\leaders\hrule height .34pt depth 0.5ex \hskip 0.5pt}
\def\downboxfill{$\moth \setbox\zo\hbox{\dtick}%
  \hskip 2pt\hbox to 0pt{$\dtick$\hss}\hrulefill \hbox to 2pt{$\dtick$\hss}$}

\def\bec{\begin{center}}
\def\ec{\end{center}}

\def\be{\begin{equation}}
\def\ee{\end{equation}}
\def\bea{\begin{eqnarray}}
\def\eea{\end{eqnarray}}
\def\ba{\begin{array}}
\def\ea{\end{array}}

\def\ft#1#2{{\textstyle{{\scriptstyle #1}
\over {\scriptstyle #2}}}}

\begin{document}

\begin{titlepage}
\begin{center}

\hfill UG-08-09 \\

\vskip 1.5 cm

{\LARGE \bf Non-linear parent action and dual gravity
\\[0.2cm]}

\vskip 1.5cm

{\bf Nicolas Boulanger\footnotemark
\footnotetext{nicolas.boulanger@sns.it; Work supported by a
``Progetto Italia'' fellowship. } and Olaf Hohm\footnotemark}
\footnotetext{o.hohm@rug.nl}

\vspace{0.8cm} $^1${\em } {\em Scuola Normale Superiore} \\
Piazza dei Cavalieri 7, 56126 Pisa, Italy\ \vspace{.5cm}

$^2${\em} {\em Centre for Theoretical Physics, University of
Groningen} \\ Nijenborgh 4, 9747 AG Groningen, The Netherlands \
\vspace{.5cm}

\vskip 0.8cm

\end{center}

\vskip 1cm

\begin{center} {\bf ABSTRACT}\\[3ex]

\begin{minipage}{13cm}
We give a reformulation of non-linear Einstein gravity, which
contains the dual graviton together with the ordinary metric and a
shift gauge field. The metric does not enter through a `kinetic'
Einstein-Hilbert term, but via topological couplings, and so the
theory does not lead to a doubling of degrees of freedom. The field
equations take the form of first-order duality relations. We analyze
the gauge symmetries and comment on their meaning with regard to the
$E_{11}$ proposal.

\end{minipage}

\vskip 3cm

\end{center}

\noindent

\vfill

June 2008
%\today

\end{titlepage}

\section{Introduction}\setcounter{equation}{0}

It is a classic result that Kaluza-Klein reduction of 11-dimensional
supergravity gives rise to exceptional hidden symmetries. Based on
this observation it has been conjectured that the
infinite-dimensional Kac-Moody algebra $E_{11}$ is a symmetry of
supergravity or possibly even M-theory \cite{West:2001as}. Part of
the evidence for this conjecture consists of the fact that the level
decompositions of $E_{11}$ with respect to the $SL(D)$ subgroups
precisely reproduce the field content of maximal supergravity in $D$
dimensions. On the supergravity side this identification requires
that one adds to each field its dual. For instance, $E_{11}$
predicts not only the 3-form of 11-dimensional supergravity, but
also a 6-form. Moreover, at higher level fields appear that
transform in mixed Young tableaux representations, and the lowest of
these can be interpreted as the dual of the metric (`dual
graviton').

At the free linearized level, Einstein gravity with metric
$h_{\mu\nu}$ can be equivalently formulated in terms of the dual
mixed Young tableaux field $C_{\mu_1\cdots\mu_{D-3}|\nu}$. To see
this, one may choose light-cone gauge and dualize one index on the
metric tensor by means of the epsilon tensor of the little group
$SO(D-2)$, resulting in the dual metric with mixed symmetries
\cite{Hull:2000zn}. Afterwards, the dual metric can be elevated to a
space-time covariant object with an associated gauge symmetry
\cite{West:2001as,West:2002jj,Hull:2001iu,Nieto:1999pn,Casini:2001gv,Bekaert:2002jn,Boulanger:2003vs,Matveev:2004ac,Ajith:2004ia},
whose covariant action has been given by Curtright
\cite{Curtright:1980yk}. However, this dualization is problematic
once the non-linear theory is considered. The no-go theorems of
\cite{Bekaert:2002uh} prove that there is no local, manifestly
Poincar\'e-invariant, non-abelian deformation of the Curtright
action, and so there is no consistent non-abelian self-interaction
of the dual graviton. One way to circumvent this no-go theorem would
be to give up space-time covariance and/or locality. In fact, if one
is willing to do so, dualization is trivially possible. One simply
has to replace inside the Einstein-Hilbert action in light-cone
gauge \cite{Goroff:1983hc}
--- which is neither local nor covariant --- the graviton by
(the Hodge-dual of) the dual graviton. A non-trivial
way would be to give up covariance, but keeping locality, as it
happens naturally in the $E_{10}$ $\sigma$-model of
\cite{Damour:2002cu}. In contrast, an essential feature of the
$E_{11}$ proposal is precisely its space-time covariance in that it
reproduces the supergravity spectra in their covariant form. So at
first sight there seems to be no way to preserve $E_{11}$ beyond the
`dual graviton barrier'.

One may still hope to avoid the no-go theorem of
\cite{Bekaert:2002uh}, which considers only pure gravity, by taking
other fields into account, as for instance 3- and 6-form of $D=11$
supergravity, or the original metric itself. The former possibility
seems to be unlikely since the Kac-Moody approach actually applies
not only to maximal supergravity, but in particular also to pure
gravity (then based on the Kac-Moody algebra $A_{D-3}^{+++}$), where
these fields are not available. The idea of adding to the action of
the dual graviton the original Einstein-Hilbert term, in order to
possibly obtain consistent cross interactions, is equally
unpromising since, even supposing the existence of such cross
interactions, it would double the degrees of freedom, in contrast to
the expectation that we should ultimately recover ordinary
(super-)gravity.

So the question we should really ask is a different one, namely
whether there exists a theory, which is
\begin{itemize}
  \item[(i)]
   classically equivalent to non-linear Einstein gravity,
  \item[(ii)]
   contains besides the metric the dual metric and
  \item[(iii)]
   is covariant and local.
 \end{itemize}
The idea of a formulation in which the metric and its dual appear
simultaneously in itself is not new. However, while sofar these
attempts abandoned space-time covariance and/or locality
\cite{Nurmagambetov:2006vz,Ellwanger:2006hy}, we will see below,
that it is surprisingly straightforward to satisfy all of the
requirements (i)--(iii). For this we will mimic an approach, which
has recently been proven to be very fruitful in the context of
gauged supergravity
\cite{Nicolai:2000sc,Nicolai:2001sv,Nicolai:2003bp,deWit:2003ja}
(see also \cite{deWit:2008ta} and references therein). Specifically,
we will start from a certain covariantisation of the Curtright
action and add a topological (Chern-Simons like) term containing the
original metric. The resulting theory is then proven to be
equivalent to Einstein gravity.

This paper is organized as follows. In Section~\ref{sec:lingrav} we
review the dualization of the graviton in the linearization, and
discuss the symmetries of the covariant action for the dual
graviton. We turn to non-linear gravity in Section~\ref{sec:toymod},
where we first explain our strategy with a toy model of
vector$\,$--$\,$scalar duality in $D=3\,$. This is then used to
derive a non-linear action, called `parent action', which
simultaneously contains the graviton and its dual. We comment on its
symmetry structure in view of the $E_{11}$ proposal and conclude in
Section~\ref{sec:parent}.

\textit{Note added after publication:}
The equations of motion (\ref{dualitygrav}) and (\ref{2ndduality})
following from our action (\ref{nonlin}) must be equivalent 
to the equations (4.12) and (4.13) given by West in \cite{West:2002jj}.
Both sets of equations give a first-order formulation of 
non-linear Einstein gravity and contain the graviton and its dual
together with an extra field (denoted by $Y$ in our section 
\ref{sec:toymod}). Our results can thus be viewed as 
complementing those of \cite{West:2002jj} by an  
action principle and, systematically following the lines of 
\cite{Nicolai:2003bp,deWit:2003ja}, by giving  
a rationale for the introduction of the extra field. 
We thank P.~West for discussions on this point.

%%%%%%%%%%%%%%%%%%%%%%%%%%%%%%%%%%%%
\section{Linearized dual gravity}\label{sec:lingrav}
%%%%%%%%%%%%%%%%%%%%%%%%%%%%%%%%%%%%%

We start by reviewing the dualization of the graviton at the
linearized level, as given, for instance, in
\cite{West:2001as,Boulanger:2003vs}. For this one uses that the
Einstein-Hilbert action based on the vielbein $e_{\mu}{}^{a}$ can be
written, up to boundary terms, as \cite{Weyl:1929fm}\footnote{We
choose the space-time signature to be $(-+\cdots +)$. The epsilon
symbol is defined by $\varepsilon^{012\cdots}=+1$,
{\textit{i.e.}}~as a density, such that
$e^{-1}\varepsilon^{\mu_1\ldots\mu_{D}}$ transforms as a tensor.}
 \bea\label{EH}
  S_{\rm EH} \ = \ -\int d^D x\hspace{0.1em}e\left(\Omega^{abc}\Omega_{abc}+
  2\Omega^{abc}\Omega_{acb}-4\Omega_{ab}{}^b\Omega^{ac}{}_{c}\right)\;,
 \eea
where
 \bea
  \Omega_{ab}{}^{c} \ = \
  e_{a}{}^{\mu}e_{b}{}^{\nu}\left(\partial_{\mu}e_{\nu}{}^{c}-\partial_{\nu}e_{\mu}{}^{c}\right)
 \eea
are the coefficients of anholonomy. This form of the
Einstein-Hilbert action can be recast into first-order form by
introducing an auxiliary field $Y_{ab|c}=-Y_{ba|c}$,
 \bea\label{first}
  S[Y,e] \ = \ -2\int d^Dx\hspace{0.1em}e
  \left(Y^{ab|c}\Omega_{abc}-\ft12Y_{ab|c}Y^{ac|b}+\tfrac{1}{2(D-2)}Y_{ab|}{}^{b}Y^{ac|}{}_{c}\right)\;.
 \eea
The field equation of $Y$ can be used to solve for it in terms of
$\Omega$,
 \bea\label{Ysol}
  Y_{ab|c} \ = \
  \Omega_{abc}-2\Omega_{c[ab]}+4\eta_{c[a}\Omega_{b]d}{}^{d}\;.
 \eea
After reinserting (\ref{Ysol}) into (\ref{first}), one precisely
recovers the Einstein-Hilbert action in the form (\ref{EH}). In
fact, the action (\ref{first}) coincides with the standard first
order action with the spin connection as independent field, up to a
mere field redefinition, which replaces the spin connection by
$Y_{ab|c}\,$. For later use we note that (\ref{first}) has the same
symmetries as the original Einstein-Hilbert action. First, it is
manifestly diffeomorphism invariant. Moreover, the invariance of the
second-order action (\ref{EH}) under the local Lorentz group can be
elevated to a symmetry of the first-order action by requiring that
the auxiliary $Y_{ab|c}$ transforms as
 \bea\label{auxsym}
  \delta_{\Lambda}Y_{ab|c} \ = \
  -2e_{c}{}^{\mu}\partial_{\mu}\Lambda_{ab}-4\eta_{c[a}e^{\mu
  d}\partial_{\mu}\Lambda_{b]d}-2\Lambda^{d}{}_{[a}Y_{b]d|c}
  +\Lambda^{d}{}_{c}Y_{ab|d}\;.
 \eea

In order to obtain the dual graviton from (\ref{first}) we have to
consider the linearized theory and vary with respect to the metric.
Before we linearize, it turns out to be convenient to first rewrite
the action in terms of the Hodge dual of $Y^{ab|c}$,
 \bea\label{dualY}
  Y^{ab|c} \ = \ \tfrac{1}{(D-2)!}\epsilon^{abc_1\cdots c_{D-2}}
  Y_{c_1\cdots c_{D-2}|}{}^{c}\;.
 \eea
This yields \bea\label{firstdual}
 \begin{split}
  S \ = \ -\tfrac{2}{(D-2)!}\int
  d^Dx\hspace{0.1em}e\Big(&\epsilon^{abc_1\ldots
  c_{D-2}}Y_{c_1\ldots c_{D-2}|}{}^{c}\Omega_{abc}+\tfrac{D-3}{2(D-2)}
  Y^{c_1\ldots c_{D-2}|b}Y_{c_1\ldots
  c_{D-2}|b}\\
  &-\tfrac{D-2}{2}Y^{c_1\ldots c_{D-3}a|}{}_{a}Y_{c_1\ldots
  c_{D-3}b|}{}^{b}+\tfrac{1}{2}Y^{c_1\ldots c_{D-3}a|b}Y_{c_1\ldots
  c_{D-3}b|a}\Big)\;.
 \end{split}
\eea In the linearisation around flat space,
$e_{\mu}{}^{a}=\delta_{\mu}{}^{a}+\kappa\, h_{\mu}{}^{a}$, we can
ignore the distinction between flat and curved indices. In
particular, we have
$\Omega_{\mu\nu\rho}=2\partial_{[\mu}h_{\nu]\rho}\,$, where the
field $h_{\mu\nu}$ has no symmetry. The field equation for
$h_{\mu\nu}$ is
 \bea\label{inte}
  \partial_{[\mu_1}Y_{\mu_2\ldots\mu_{D-1}]|\nu} \ = \ 0\;.
 \eea
The Poincar\'e lemma then implies that $Y$ is the curl of a
potential $C_{\mu_1\ldots\mu_{D-3}|\nu}$ (the `dual graviton'),
which is completely antisymmetric in its first $D-3$ indices,
 \bea\label{intesol}
  Y_{\mu_1\ldots\mu_{D-2}|\nu} \ = \
  \partial_{[\mu_1}C_{\mu_2\ldots\mu_{D-2}]|\nu}\;.
 \eea
Inserting this back into (\ref{firstdual}) yields a consistent
action $S[C]$ for the dual graviton.

Up to now, $C_{\mu_1\ldots\mu_{D-3}|\nu}$ as defined by
(\ref{intesol}) does not transform in an irreducible $GL(D)$
representation since also $Y$ does not possess a specific
Young-diagram symmetry. However, one may
check~\cite{Boulanger:2003vs} that, after inserting (\ref{intesol})
into the linearisation of (\ref{firstdual}), the resulting action
$S[C]$ is invariant under the following St\"uckelberg symmetry
 \bea\label{stuckel}
  \delta_{\Lambda}C_{\mu_1\ldots\mu_{D-3}|\nu} \ = \
  -\Lambda_{\mu_1\ldots\mu_{D-3}\nu}\;,
 \eea
with completely antisymmetric shift parameter. Therefore, the
totally antisymmetric part of $C_{\mu_1\ldots\mu_{D-3}|\nu}$ can be
gauge-fixed to zero inside $S[C]\,$, giving rise to the dual
graviton with a $(D-3,1)$ Young-diagram symmetry.\footnote{In this
paper we denote by $(p,q)$ two-column Young diagrams in the
antisymmetric basis with $p$ boxes in the first column and $q$ boxes
in the second column.} In other words, in the action $S[C]$ the dual
graviton appears in the so-called frame-like formulation. The latter
is the analogue of the vielbein formalism, in which the linearized
Lorentz transformations act as St\"uckelberg transformations, and
which can be generalized to arbitrary-spin fields
\cite{Vasiliev:1980as,Vasiliev:1986td,Lopatin:1987hz} (more
recently, see also
\cite{Zinoviev:2003ix,Alkalaev:2003qv,Engquist:2007yk,Skvortsov:2008vs}).

Let us stress that even though (\ref{first}) and thus
(\ref{firstdual}) are first-order formulations of
\textit{non-linear} Einstein gravity, the identification of the dual
graviton in (\ref{intesol}) is only possible in the linearisation,
since in the full theory the integrability condition (\ref{inte}) is
violated \cite{West:2001as}. This is in agreement with the fact
that there is no non-abelian self-interacting theory for the dual
graviton \cite{Bekaert:2002uh}.

Before we proceed, let us examine the free theory of the dual
graviton in more detail. In order to indicate that the field now
carries a specific Young-diagram symmetry, we denote it by
$D_{\mu_1\ldots\mu_{D-3}|\nu}$. The characteristics of those mixed
Young tableaux fields have been studied independently in
\cite{Curtright:1980yk,Aulakh:1986cb}. First of all, it transforms
under two types of gauge transformations,
 \bea\label{dualgauge}
  \delta
  D_{\mu_1\cdots\mu_{D-3}|\nu}=
   \partial_{[\mu_1}\alpha_{\mu_2\cdots\mu_{D-3}]|\nu}
  +\partial_{[\mu_1}\beta_{\mu_2\cdots\mu_{D-3}]\nu}
   -(-1)^{D-3}\partial_{\nu}\beta_{\mu_1\cdots\mu_{D-3}}\;.
 \eea
Here, $\alpha$ possesses the $(D-4,1)$ Young-diagram symmetry, and
$\beta$ is completely antisymmetric. Consequently, (\ref{dualgauge})
is consistent with the Young tableau symmetry of
$D_{\mu_1\ldots\mu_{D-3}|\nu}$. The $\beta$-transformations are the
`dual' diffeomorphisms. For instance, in $D=4$, where the metric is
self-dual, (\ref{dualgauge}) reads
$\delta_{\beta}D_{\mu\nu}=\partial_{\mu}\beta_{\nu}+\partial_{\nu}\beta_{\mu}$.
In analogy to the ordinary graviton, there is no invariant field
strength which is first order in derivatives, but only a second
order Riemann tensor-like object. However, for the
$\alpha$-transformations an invariant field strength is simply given
by
 \bea\label{fieldstrength}
  F_{\mu_1\cdots\mu_{D-2}|\nu} \ = \
  \partial_{[\mu_1}D_{\mu_2\cdots\mu_{D-2}]|\nu}\;.
 \eea
An invariant action (the Curtright action) can then be written as
$S[C]=\int d^Dx\, {\cal L}_{\rm C}(F)\,$, where
\begin{eqnarray}\label{Caction}
\begin{split}
  {\cal L}_{\rm C}(F) \; = & \; \frac{D-3}{2(D-2)} \,F^{\mu_1\cdots\mu_{D-2}|\nu}F_{\mu_1\cdots\mu_{D-2}|\nu}
  -\ft12(D-2)\, F^{\mu_1\cdots\mu_{D-3}\rho|}{}_{\rho}
  F_{\mu_1\cdots\mu_{D-3}\lambda|}{}^{\lambda} \\
   & \hspace*{.3cm} + \tfrac{1}{2}F^{\mu_1\cdots\mu_{D-3}\nu|\rho}
  F_{\mu_1\cdots\mu_{D-3}\rho|\nu}\;.
\end{split}
\end{eqnarray}
Here the coefficients are fixed by requiring gauge-invariance under
$\beta$-transformations. Up to a global pre-factor, this is
precisely the action one obtains by inserting (\ref{intesol}) into
(\ref{firstdual}). And, in fact, the distinction between $C$ and $D$
becomes redundant, since due to the symmetry (\ref{stuckel}), in the
action the antisymmetric part of $C$ drops out. To be more precise,
the Lagrangian ${\cal L}_{\rm C}(F)$ given above is invariant under
(\ref{stuckel}) up to a total derivative.

%%%%%%%%%%%%%%%%%%%%%%%%%%%%%%%%%%%%%%%%%%%%%%%%%%%%%%%%%
\section{Covariant theory of non-linear dual gravity}\label{sec:toymod}
%%%%%%%%%%%%%%%%%%%%%%%%%%%%%%%%%%%%%%%%%%%%%%%%%%%%%%%%%%%%

In this section we are going to propose a non-linear theory
featuring the dual graviton, which still contains the original
metric via a topological term. The resulting theory will be
equivalent to ordinary general relativity. In order to motivate our
approach, we first recall a non-trivial duality for non-abelian
gauge vectors encountered in gauged supergravity.

\subsection{A toy model: Dualizing non-abelian vectors}\label{toy}

As is well known, in $D=3$ a free theory of abelian Maxwell vectors
is dual to a free theory of massless scalars. However, once the
gauge vectors are promoted to non-abelian Yang-Mills gauge fields,
or if they are coupled to charged matter, this duality breaks down.
As has been shown in \cite{Nicolai:2003bp,deWit:2003ja}, it is
nevertheless possible to assign all propagating degrees of freedom
to scalar fields, while the gauge vectors appear only through
topological Chern-Simons terms. In other words, besides the dual
scalars the action still contains the (non-abelian) gauge vectors.

To illustrate this, let us start directly from the non-linear
action, whose corresponding Lagrangian is given by
 \bea\label{gaction}
  %S[\varphi,A,B]=\int d^3x\,{\cal L}_{g}(\varphi,A,B)\;,\quad
  {\cal L}_{g}(\varphi,A,B) \ = \ -\frac{1}{2}\left(\kappa^{ab}
  {\cal D}_{\mu}\varphi_a{\cal D}^{\mu}\varphi_b -
  \varepsilon^{\mu\nu\rho}B_{\mu a}{\cal{F}}_{\nu\rho}^{a}\right)\;,
 \eea
which depends on scalars $\varphi_a$ and gauge vectors
$A_{\mu}^{a}$, $B_{\mu a}\,$. Here the covariant derivatives and
non-abelian field strengths are defined by
 \begin{eqnarray}
  {\cal D}_{\mu}\varphi_a &=& \partial_{\mu}\varphi_a +
  gf_{ab}{}^{{}c}A_{\mu}^b\varphi_c+B_{\mu a}\;, \\
  {\cal{F}}_{\mu\nu}^{a} &=&
  \partial_{\mu}A_{\nu}^{a}-\partial_{\nu}A_{\mu}^{a}
  +g f_{bc}{}^{a}A_{\mu}^{b}A_{\nu}^{c}\;,
 \end{eqnarray}
where $f_{ab}{}^{c}$ are the structure constants of a compact semi-simple
real Lie algebra with invariant Cartan-Killing form
$\kappa^{ab}\propto\,\delta^{ab}\,$.
Therefore, (\ref{gaction}) is manifestly invariant under the gauge symmetries
 \begin{eqnarray}\label{gaugedsym}
  \delta\varphi_a &=& -\Sigma_a -gf_{ab}{}^c\epsilon^b\varphi_c\;,
 \label{gaugevarphi}\\
  \delta A_{\mu}^a &=& \partial_{\mu}\epsilon^a
  +gf_{bc}{}^{a}A_{\mu}^b\epsilon^c\;,
 \label{gaugevarA}\\
  \delta B_{\mu a} &=& \partial_{\mu}\Sigma_a
  +gf_{ab}{}^cA_{\mu}^b\Sigma_c-gf_{ab}{}^c\epsilon^bB_{\mu c}\;.
 \label{gaugevarB}
\end{eqnarray}
Even though this theory describes charged scalars and non-abelian
gaugings, it is still possible to dualize the scalars to vectors. To
see this, we observe that due to the presence of a Chern-Simons
term, the field equations of the gauge vectors are duality relations
between vectors and scalars. Specifically, varying with respect to
$B_{\mu a}$ gives
 \bea
  {\cal D}^{\mu}\varphi_a \ = \
  \frac{1}{2}\kappa_{ab}\varepsilon^{\mu\nu\rho}{\cal{F}}_{\nu\rho}^{b}\;.
 \eea
This can be used to solve for $B_{\mu a}$ in terms of $A_{\mu}^{a}$
and $\varphi_a\,$. After reinsertion into (\ref{gaction}), one
recovers precisely the non-abelian Yang-Mills Lagrangian,
 \bea\label{YMaction}
  {\cal L}_{g}(A) \ = \ -\frac{1}{4} \kappa_{ab}
  {\cal{F}}^{\mu\nu a}{\cal{F}}_{\mu\nu}^{b}\;.
 \eea
(More conveniently, one may first use the shift symmetry spanned by
$\Sigma_a$ in order to gauge-fix $\varphi_a$ to zero. Then,
on-shell, $B_{\mu a}$ is entirely expressed in terms of
$A_{\mu}^{a}\,$.)

In the ungauged limit $g\rightarrow 0$, the covariant derivatives
reduce to mere St\"uckelberg derivatives,
$D_{\mu}\varphi_a=\partial_{\mu}\varphi_a+B_{\mu a}$, while the
Chern-Simons term becomes abelian. In this limit the symmetries
reduce to the abelian
 \bea\label{abelian}
  \delta\varphi_a \ = \ -\Sigma_a\;, \qquad
  \delta B_{\mu a} \ = \ \partial_{\mu}\Sigma_a\;, \qquad
  \delta A_{\mu}^{a} \ = \ \partial_{\mu}\epsilon^{a}\;,
 \eea
and integrating out $B_{\mu a}$ results into the (positive) sum of
Maxwell actions, of which (\ref{YMaction}) provides a consistent
non-linear deformation.\footnote{In gauged supergravity it is
usually convenient to have a different dependence on the gauge
coupling $g$, which is such that the Chern-Simons term vanishes for
$g\rightarrow 0$ \cite{Nicolai:2000sc}. The chosen assignment of the
deformation parameter here is necessary in order to have the same
`duality-covariant' form in the ungauged theory as well.}

Let us finally analyze the deformation of the gauge symmetries in
more detail. At first sight, the gauging deforms the abelian gauge
transformations (\ref{abelian}) for the $\epsilon^{a}$ as well as
for the $\Sigma_a$ in that the latter transform non-trivially under
the former (see eq.~(\ref{gaugedsym})). In fact, the gauge
transformations close according to
 \bea\label{semidirect}
  \big[ \delta_{\epsilon},\delta_{\Sigma}\big] \ = \
  \delta_{\tilde{\Sigma}}\;, \qquad
  \tilde{\Sigma}_a \ = \ gf_{ab}{}^{c}\epsilon^{b}\Sigma_{c}\;,
 \eea
indicating a semi-direct product between the Yang-Mills gauge group
and the translations. However, it is possible to show that the only
true deformation of the gauge algebra concerns the Yang-Mills
transformations spanned by $\epsilon^{a}$. More precisely, one can
redefine the parameters and the fields in such a way that the
seemingly semi-direct product (\ref{semidirect}) trivializes,
leaving separate Yang-Mills transformations and abelian
translations. To show this we redefine the shift parameter according
to
 \bea
  \bar{\Sigma}_a \ = \
  \Sigma_{a}+gf_{ab}{}^{c}\epsilon^{b}\varphi_{c}\;,
 \eea
and the gauge field $B_{\mu a}$ by
 \bea
  \bar{B}_{\mu a} \ = \ B_{\mu a} +
  gf_{ab}{}^{c}A_{\mu}^{b}\varphi_c\;.
 \eea
After this redefinition, in total the fields transform as
 \bea
  \delta\varphi_a \ = \ -\bar{\Sigma}_a\;, \qquad
  \delta\bar{B}_{\mu a} \ = \
  \partial_{\mu}\bar{\Sigma}_a-gf_{ab}{}^{c}\epsilon^{b}
   {\cal D}_{\mu}\varphi_{c}\;.
 \eea
In other words, the gauge transformations on $\varphi_a$ and $B_{\mu
a}$ are as in the free case (\ref{abelian}), up to a correction by
the gauge-covariant derivative ${\cal D}_{\mu}\varphi_{a}$. However,
as the latter is shift-invariant, one finds that the commutator
(\ref{semidirect}) indeed trivializes,
$[\delta_{\epsilon},\delta_{\bar{\Sigma}}] = 0$.

Before we proceed with the dual graviton, let us briefly comment on
the properties of this theory in view of the $E_{11}$ proposal. The
reader might be disturbed by the fact that the enhancement of
symmetries has been achieved through the introduction of a simple
shift invariance, expressing a trivial product structure. However,
this is in precise correspondence to what happens in the relation
between $E_{11}$ and ordinary $p$-form gauge symmetries
\cite{Bergshoeff:2007vb}. For instance, a 2-form is taken to
transform as $\delta B_{\mu\nu}=\partial_{[\mu}\Lambda A_{\nu]}$,
for which the algebra closes according to the ($p$-form truncation
of the) $E_{11}$ algebra. This transformation can in turn be
redefined such that $\delta B_{\mu\nu}=-\Lambda F_{\mu\nu}$, with
the gauge-invariant field strength $F_{\mu\nu}$. Therefore, the
commutator vanishes, hence trivializing the algebra. Given these
similarities, we apply the presented scheme of `non-abelian
dualization' to the dual graviton and comment on the
supergravity/Kac-Moody correspondence later on.

\subsection{Linear dual gravity and its symmetries}

In the last section we have seen that in $D=3\,$ even the
non-abelian, that is, self-interacting Yang-Mills theory, can be
dualized to a scalar theory, which then contains both the field and
its dual. Consequently, this amounts to an enhancement of the gauge
symmetry, since the action (\ref{gaction}) exhibits besides the
standard Yang-Mills symmetry additional local symmetries spanned by
$\Sigma_a$ (even though, as we have seen, their product structure is
trivial). As we have argued in the introduction, we expect something
similar for gravity. By strict analogy, we are looking for a
non-linear and covariant theory with kinetic terms for the dual
graviton, but which still contains topological terms for the
original graviton.

Let us start with the free theory in frame-like formulation, with
kinetic terms for the dual graviton $C^{\qquad\quad
~a}_{\mu_1\ldots\mu_{D-3}}\,$. In addition, we introduce a
St\"uckelberg gauge field $Y^{\qquad\quad
~a}_{\mu_1\cdots\mu_{D-2}}\,$ and a shift-invariant form
$\hat{F}^{\qquad\quad ~a}_{\mu_1\cdots\mu_{D-2}}$ of the field
strength $F_{\mu_1\cdots\mu_{D-2}}{}^a \ = \
  \partial_{[\mu_1}C_{\mu_2\cdots\mu_{D-2}]}{}^a\,$,
 \bea\label{shiftfield}
  \hat{F}^{\qquad\quad ~a}_{\mu_1\cdots\mu_{D-2}} \ = \
  F^{\qquad\quad ~a}_{\mu_1\cdots\mu_{D-2}} +
  Y^{\qquad\quad ~a}_{\mu_1\cdots\mu_{D-2}}\;.
 \eea
The field strength $\hat F$ is invariant under
 \bea\label{shiftsym}
  \delta Y^{\qquad\quad ~a}_{\mu_1\cdots\mu_{D-2}} \ = \
  \partial^{}_{[\mu_1}\Sigma^{\qquad\quad \;~a}_{\mu_2\cdots\mu_{D-2}]}\;,
  \qquad
  \delta C^{\qquad\quad ~a}_{\mu_1\cdots\mu_{D-3}} \ = \
  -\Sigma^{\qquad\quad ~a}_{\mu_1\cdots\mu_{D-3}}\;.
 \eea
In order to make the transition to the non-linear theory in the next
section more transparent, we have kept the formal distinction
between flat and curved indices, which are related by the trivial
background vielbein ${\bar e_{\mu}^{~\,a}}=\delta_{\mu}{}^a$. We
recall that the vierbein is expanded, around flat spacetime, as
$e_{\mu}^{~\,a}={\bar e_{\mu}^{~\,a}}+\kappa\,h^{~\,a}_{\mu}\,$.
Here we have taken all fields to be in reducible representations,
i.e., the fields $C$ and $Y$ as well as the transformation parameter
$\Sigma$ possess an antisymmetric part, after converting all the
indices into curved indices. In total, we consider the action
 \bea\label{dualgrav}
  S=\int d^Dx\;{\cal L}(h,C,Y)\;,\quad
   {\cal L}(h,C,Y) \ = \ {\cal L}_{\rm C}(\hat{F}) +
  2\,\varepsilon^{\mu_1\cdots\mu_{D-2}\nu\rho}\;
  Y_{\mu_1\cdots\mu_{D-2}}{}^{a}\;\partial_{\nu}h_{\rho a}\;,
 \eea
where we added in complete analogy to (\ref{gaction}) a topological
term containing the ordinary graviton $h^{~\,a}_{\mu}\,$. Let us
stress that here also $h_{\mu\nu}$ is not in an irreducible Young
tableau, but carries an antisymmetric part.

The physical content of (\ref{dualgrav}) can be analyzed as follows.
Varying with respect to $h^{~\,a}_{\mu}$ yields
$\partial_{[\mu_1}Y_{\mu_2\ldots\mu_{D-1}]}{}^{a}=0\,$,
{\textit{i.e.}}~the shift gauge field is pure gauge and can
therefore be gauged to zero by virtue of (\ref{shiftsym}). The
action for the remaining field $C_{\mu_1\ldots\mu_{D-3}}{}^a$ is
then precisely the Curtright action for the dual graviton. On the
other hand, varying with respect to $Y$ one obtains a `duality
relation' between $h$ and $\hat{F}\,$. Integrating out $Y$ yields
the linearized action for gravity, where the antisymmetric part of
$h_{\mu\nu}$ appears in the corresponding Lagrangian only through
total derivatives. This is essentially the same calculation as the
one which led from the first-order, quadratic action
(\ref{firstdual}) back to the quadratic part of the Einstein-Hilbert
action (\ref{EH}), the only difference being the presence of $C$ in
the field strength (\ref{shiftfield}). However, the latter cancels
out, as it should be due to the shift invariance (\ref{shiftsym}).
To summarize, the parent action based on (\ref{dualgrav}) contains
both the graviton and its dual and consistently describes the free
dynamics of either of them.

Let us briefly analyze the symmetries of the free theory
(\ref{dualgrav}), apart from the manifest shift symmetry
(\ref{shiftsym}). The diffeomorphisms and local Lorentz
transformations on $h_{\mu}{}^a$ read
 \bea\label{diff}
  \delta h_{\mu}{}^a \ = \
  \partial_{\mu}\xi^a - \Lambda^a{}_{\mu}\;,
 \eea
while all other fields are invariant under $\xi^a\,$. The dual
diffeomorphisms and $\alpha$-transformations `unify' to one
symmetry, given by
 \bea\label{dualdiff}
  \delta_{\gamma}C_{\mu_1\ldots\mu_{D-3}}{}^a \ = \
  \partial_{[\mu_1}\gamma_{\mu_2\ldots\mu_{D-3}]}{}^a\;.
 \eea
More precisely, $\gamma$ carries the Young-diagram symmetries
 \bea
  (D-4)\; \otimes\; (1) \ = \ (D-4,1) \; \oplus \; (D-3)\;,
 \eea
whose irreducible parts are identified with $\alpha$ and $\beta$,
respectively. That both symmetries are manifest is due to the
frame-like formulation. In fact, instead of the dual diffeomorphisms
it is now the local Lorentz symmetry which acts non-trivially and
fixes the relative coefficients in ${\cal L}(h,C,Y)\,$. It reads
\bea \label{linLorY}
  \delta^{(0)}_{\Lambda} Y_{\mu_1\cdots\mu_{D-2}|a} & =&
  \partial_{[\mu_1}
  \big({\bar e}_{\mu_2}{}^{b_2}\ldots {\bar e}_{\mu_{D-2}]}{}^{b_{D-2}}
   \;\tilde\Lambda_{b_2\cdots b_{D-2}a}\big)\;,
  \\
  \delta^{(0)}_{\Lambda} C_{\mu_1\cdots\mu_{D-3}|a} & = &
  {\bar e}_{\mu_1}{}^{b_1}\ldots {\bar e}_{\mu_{D-3}}{}^{b_{D-3}}
   \;\tilde\Lambda_{b_1\cdots b_{D-3}a} \;,
\label{linLorC}
 \eea
where $\tilde{\Lambda}$ is proportional to the Hodge dual of
$\Lambda\,$,
 \bea\label{tildeL}
  \tilde{\Lambda}_{a_1\ldots a_{D-2}} \ = \
  \ft12(-1)^{D-3}(D-2)\,
 \epsilon_{a_1\ldots a_{D-2}bc}\;\Lambda^{bc}\;.
 \eea
Thus, the Lorentz parameter can be used to gauge away either the
antisymmetric part of the metric or of its dual (but not
simultaneously). Such gauge-fixing requires compensating gauge
transformations for the symmetries (\ref{diff}) and
(\ref{dualdiff}), which in turn reintroduces the non-manifest
invariance of the action either under the diffeomorphisms
$\delta_{\xi}h_{\mu\nu}=\partial_{\mu}\xi_{\nu}+\partial_{\nu}\xi_{\mu}$
or under their dual (\ref{dualgauge}).

%--------------------------------------------------------
\subsection{Non-linear dual gravity}\label{nonlingrav}
%--------------------------------------------------------

We turn now to the non-linear theory. We proceed again in analogy to
the vector-scalar example (\ref{gaction}), where the step from the
linear to the non-linear theory was simply given by covariantising
the field strengths and derivatives with respect to the Yang-Mills
gauge group. Thus, here we are going to make the action invariant
under the full diffeomorphism group by introducing the dynamical
metric in the kinetic terms for the dual graviton.

The action reads
 \bea\label{nonlin}
  S[e,C,Y] &=& \int d^Dx \big[ {\cal L}_C(e,\hat{F})
   +2\kappa^{-1}\,\varepsilon^{\mu_1\ldots\mu_{D-2}\nu\rho}\,
  Y_{\mu_1\ldots\mu_{D-2}|a}\;\partial_{\nu}e_{\rho}{}^{a} \big]\;,
 \eea
where we introduced the `covariantized' Curtright Lagrangian
 \begin{eqnarray}
   {\cal L}_C(e,\hat{F}) &=& \tfrac{D-3}{2(D-2)}\,e\,
   \hat{F}^{\mu_1\ldots\mu_{D-2}|a}\hat{F}_{\mu_1\ldots\mu_{D-2}|a}
   -\tfrac{D-2}{2}\;e\, e_{\nu}{}^{a}\,e_{b}{}^{\rho}\,
   \hat{F}^{\mu_1\ldots\mu_{D-3}\nu|}{}_{a}\,
   \hat{F}_{\mu_1\ldots\mu_{D-3}\rho|}{}^{b}
    \nonumber \\
   && +\tfrac{1}{2}\,e\,e_{\nu}{}^{b}\,e_{a}{}^{\rho}\,
   \hat{F}^{\mu_1\ldots\mu_{D-3}\nu|a}\,\hat{F}_{\mu_1\ldots\mu_{D-3}\rho|b}
   \;.
  \nonumber
 \end{eqnarray}
Here, all curved indices are raised and lowered with the metric
$g_{\mu\nu}=e_{\mu}{}^{a}\,e_{\nu}{}^{b}\,\eta_{ab}$ derived from
$e_{\mu}{}^{a}={\bar e}_{\mu}{}^{a}+\kappa\,h_{\mu}{}^{a}\,$, and we
introduced $\kappa$ in the topological term, such that we recover
the free Lagrangian (\ref{dualgrav}) in the limit $\kappa\rightarrow
0$. The shift-invariant field strength is not modified and still
given by (\ref{shiftfield}). Due to the appearance of inverse
vielbeins and the determinant $e$ this action is indeed a non-linear
deformation (in $\kappa$) of (\ref{dualgrav}). The action
(\ref{nonlin}) is equivalent to the non-linear Einstein-Hilbert
action, which can be reobtained by integrating out $Y$. In fact,
this can be made completely manifest by gauge-fixing the shift
symmetry such that $C=0$ and then converting all indices into flat
ones. The resulting action then coincides with the first-order form
(\ref{firstdual}).

Let us now turn to the non-linear symmetries of (\ref{nonlin}).
First, it is manifestly diffeomorphism invariant due the presence of
a dynamical metric (and for the topological term anyway). In
particular, due to the frame-like formulation, we do not need to
introduce Christoffel connections, since the (curved) space-time
indices are totally antisymmetric. The dual diffeomorphisms together
with the $\alpha$-transformations (both parameterized by
$\gamma^{a}$) act as in the linearized theory according to
(\ref{dualdiff}), leaving the field strength (\ref{shiftfield})
manifestly invariant. The shift symmetries are still given by
(\ref{shiftsym}).

The only non-trivial symmetry is the Lorentz symmetry, which we
assume to act in the standard way on the vielbein,
 \bea\label{ordlor}
  \delta_{\Lambda}e_{\mu}{}^{a} \ = \
  -\Lambda^{a}{}_{b}\;e_{\mu}{}^{b}\;.
 \eea
This is only a symmetry if suitable transformations are assigned to
$C$ and $Y\,$. On the $C$ field we take the direct non-linear
covariantization of (\ref{linLorC}):
 \begin{eqnarray}\label{LorC}
   \delta_{\Lambda} C_{\mu_1\cdots\mu_{D-3}|a} &=&
   \tilde\Lambda_{\mu_1\cdots \mu_{D-3}\,a}
   -\Lambda_{ab}\; C_{\mu_1\ldots\mu_{D-3}}{}^{b}\;\quad ,
 \end{eqnarray}
with the dual Lorentz parameter
 \begin{eqnarray}
   \tilde\Lambda_{\mu_1\cdots \mu_{D-3}a} &=&
   {e}_{\mu_1}{}^{b_1}\ldots {e}_{\mu_{D-3}}{}^{b_{D-3}}
   \tilde\Lambda_{b_1\cdots b_{D-3}\,a}
 \end{eqnarray}
introduced in (\ref{tildeL}). In the gauge-fixed formulation where
$C=0$, the corresponding variation for $Y$ can simply be determined
by applying (\ref{auxsym}) to (\ref{dualY}). Then, in the full
theory, a correction term containing $C$ has to be added. In total,
we find the non-linear transformations
 \begin{eqnarray}\label{Ylor}
  \delta_{\Lambda}Y_{\mu_1\ldots\mu_{D-2}}{}^{a} &=&
  \partial_{[\mu_1}\tilde{\Lambda}_{\mu_2\ldots\mu_{D-2}]}{}^{a}
  -(D-3)\,\Omega_{[\mu_1\mu_2}{}^{\rho}
  \tilde{\Lambda}_{|\rho|\mu_3\ldots\mu_{D-2}]}{}^{a}
  \nonumber \\
  && -\Lambda^{a}{}_{b}\;Y_{\mu_1\ldots\mu_{D-2}}{}^{b}
  + (-1)^{D-3}C_{[\mu_1\ldots\mu_{D-3}}{}^{b}\;\partial_{\mu_{D-2}]}\Lambda^{a}{}_b\;.
\label{fullgt}
\end{eqnarray}
Let us note that invariance of the action (\ref{nonlin}) under these
Lorentz transformations can be most easily checked in flat indices,
for which the correction term in (\ref{Ylor}) proportional to
$\Omega_{\mu\nu}{}^{\rho}$ is not required. Actually, the role of
the second and fourth terms in (\ref{fullgt}) is to make the total
gauge transformation of the shift-invariant field strength $\hat F$
simple:
\begin{eqnarray}
(\delta_{\Lambda}+\delta_{\gamma})
\hat{F}_{\mu_1\ldots\mu_{D-2}}{}^{a}
 &=& -\Lambda^{a}{}_{b}\;\hat{F}_{\mu_1\ldots\mu_{D-2}}{}^{b} \\
 \nonumber
   &&+ \,2\,(-1)^{D-3}\,{e}_{[\mu_1}{}^{b_1}\ldots \,{e}_{\mu_{D-3}}{}^{b_{D-3}}
   \;\partial_{\mu_{D-2}]}\tilde\Lambda_{b_1\cdots b_{D-3}}{}^a
   \;.
\end{eqnarray}

The gauge transformations take a somewhat unconventional form, as
for instance the presence of the dual Lorentz parameter
(\ref{tildeL}). Moreover, the partial derivative on $\gamma^{a}$ in
(\ref{dualdiff}) is not Lorentz covariant, and so at first sight the
dual diffeomorphisms will not close with the local Lorentz group.
However, it turns out that closure is ensured by virtue of the
additional local shift symmetry in that
 \bea\label{trivalg}
  \big[ \delta_{\gamma},\delta_{\Lambda}\big] C_{\mu_1\ldots\mu_{D-3}}{}^{a}
  \ = \ \delta_{\Sigma}C_{\mu_1\ldots\mu_{D-3}}{}^{a}\;, \qquad
  \Sigma_{\mu_1\ldots \mu_{D-3}}{}^{a} \ = \ \Lambda^{a}{}_{b}
  \partial_{[\mu_1}\gamma_{\mu_2\ldots\mu_{D-3}]}{}^{b}\;,
 \eea
and similarly on $Y$. Moreover, one finds off-shell closure for the
local Lorentz group itself,
 \bea
  \big[ \delta_{\Lambda_{1}},\delta_{\Lambda_2}\big] \ = \
  \delta_{[\Lambda_1,\Lambda_2]}\;, \qquad
  \big[ \Lambda_1,\Lambda_2\big]^{ab} \ = \
  \Lambda_{1}{}^{a}{}_{c}\Lambda_{2}{}^{cb}-\Lambda_{2}{}^{a}{}_{c}\Lambda_{1}{}^{cb}\;.
 \eea
In order to verify this, it is again more convenient to work in flat
indices or, otherwise, to keep in mind that the definition of the
parameter $\tilde{\Lambda}$ in (\ref{Ylor}) involves the vielbein.

Let us now turn to the equations of motion, specifically to the
duality relation between the metric and its dual. As in the toy
model discussed in sec.~\ref{toy}, by virtue of the topological term
in (\ref{nonlin}), the duality relation follows from the action by
varying with respect to the gauge field $Y$. One finds
 \begin{eqnarray}\nonumber
  e^{-1}\varepsilon^{\mu_1\ldots\mu_{D-2}\nu\rho}\Omega_{\nu\rho}{}^{a}
  &=& -\frac{D-3}{D-2}\hat{F}^{\mu_1\ldots\mu_{D-2}|a} +(-1)^{D-3}
  (D-2)e_{\rho b}e^{a[\mu_1}\hat{F}^{\mu_2\ldots\mu_{D-2}]\rho|b} \\
  %\nonumber
  &&-(-1)^{D-3}e_{\rho}{}^{a}e_{b}{}^{[\mu_1}\hat{F}^{\mu_2\ldots\mu_{D-3}]\rho|b}\;.
 \label{dualitygrav}
 \end{eqnarray}
As a consistency check one may now verify that this non-linear
duality relation is completely gauge covariant. In particular, due
to the presence of $Y$, it transforms covariantly under the local
Lorentz group. The field equations for $C$ can be obtained from
(\ref{dualitygrav}) by acting with a derivative. In order to obtain
the Einstein equation, we have to use the field equation for
$e_{\mu}{}^{a}$, which also takes a first-order form,
 \bea\label{2ndduality}
  e^{-1}\varepsilon^{\mu\mu_1\ldots\mu_{D-1}}\partial_{\mu_1}
  Y_{\mu_2\ldots\mu_{D-1}|a}
  \ = \ \ft12 \,e^{-1}\frac{\delta{\cal L}_{\rm C}(e,\hat{F})}
  {\delta e_{\mu}{}^{a}}\;.
 \eea
In this sense, the full set of field equations --- and so the
non-linear Einstein equations --- can be written as first-order
duality relations. %\footnote{The nonlinear Lagrangian equations 
%(\ref{dualitygrav}) and (\ref{2ndduality}) 
%should be equivalent to the equations (4.12) and (4.13)
%of \cite{West:2002jj} because both sets of equations contain the same types of fields 
%and are equivalent to Einstein's equations. We are grateful to P.~West for 
%pointing this out to us.} 
 Moreover, it follows that even in presence of the
dual graviton arbitrary matter couplings can be introduced, simply
by adding to (\ref{nonlin}) the matter action. This, in fact, leaves
the first duality relation unchanged, but adds to the second duality
relation (\ref{2ndduality}) the standard energy-momentum tensor
$T^{\mu}{}_{a}\sim\delta{\cal L}_{\rm M}/\delta e_{\mu}{}^{a}$,
which in turn appears in the Einstein equation in the usual way.
Equivalently, since the shift-gauge field $Y$ will not contribute to
possible matter terms added to (\ref{nonlin}), it can be integrated
out as before, leading to the Einstein-Hilbert action augmented by
these matter couplings. This circumvents the negative findings of
\cite{Bergshoeff:2008vc}, where it has been shown that in presence
of matter the elimination of the graviton in favor of its dual is
problematic even if gravity is treated linearly.

\subsection{Symmetries and their deformation}

In this section we would like to discuss to what extent the gauge
symmetries of the non-linear theory (\ref{nonlin})
represent deformations of the symmetries of the free theory (\ref{dualgrav}).
Analogously to what happens in the vector-scalar example presented before,
we expect that the nonlinear action can be obtained
from the free one by a deformation that does not affect the gauge algebra,
apart from diffeomorphisms and Lorentz transformations.

To see this, let us recover (\ref{fullgt}) from a different
perspective. First we deform the free Lagrangian and the corresponding
abelian gauge transformations in such a way that
\begin{eqnarray}
\delta^{(0)}_{\Lambda} Y_{\mu_1\ldots\mu_{D-2}|a} \longrightarrow
\delta^{(0)}_{\Lambda}
Y_{\mu_1\ldots\mu_{D-2}|a} + \hat{F}_{\mu_1\ldots\mu_{D-2}|b}\;
\Lambda^b_{~a}\;.
\end{eqnarray}
This deformation does not change the gauge algebra involving
$\Sigma$ and $\gamma$ due to the shift invariance of $\hat F\,$.
Then, we redefine the gauge parameter $\Sigma_{\mu_1\ldots\mu_{D-3}|a}$ by
\begin{eqnarray}
\Sigma_{\mu_1\ldots\mu_{D-3}|a}\longrightarrow
\Sigma_{\mu_1\ldots\mu_{D-3}|a} - C_{\mu_1\ldots\mu_{D-3}|b}\;\Lambda^b_{~a}\;.
\end{eqnarray}
This procedure generates the algebra (\ref{trivalg}) and gives the
gauge transformation (\ref{fullgt}), apart from the second term
therein, that reflects the Lorentz and diffeomorphism deformations.

Finally, let us briefly comment on the connection between the
discussed symmetries of the dual graviton theory and the hidden
symmetries found in dimensional reductions. Since the appearance of
the latter symmetries relies on the dualization of certain fields,
one might expect that, after introducing the dual graviton, they are
at least partially present already in the higher-dimensional theory.
For instance, in the reduction of pure gravity from $D=4$ to $D=3$ a
hidden $SL(2,\mathbb{R})$ appears, which acts non-linearly on
scalars $\phi$ and $\varphi$, which are the dilaton arising from the
metric and the dual of the Kaluza-Klein vector. (For a review see,
e.g., \cite{Nicolai:1991tt}.) Specifically, among the
$SL(2,\mathbb{R})$ generators $h$, $e$ and $f$ in the standard
Chevalley basis, $h$ originates from the higher-dimensional
diffeomorphism invariance and acts linearly, while $e$ and $f$
correspond to non-linear symmetries \cite{Nicolai:1991tt},
 \bea
  \delta_{\lambda} (e)\varphi \ = \ \lambda\;, \qquad
  \delta_{\alpha} (f)\phi \ = \ 2\alpha\phi\varphi\;, \quad
  \delta_{\alpha} (f)\varphi \ = \
  \alpha\left(\varphi^2-\phi^2\right)\;.
 \eea
In the reformulation given in sec.~\ref{nonlingrav}, there are
additional Kaluza-Klein components originating from the dual
graviton $C_{\mu}{}^{a}$, whose `dilaton' component $C_{3}{}^{3}$
one might identify with $\varphi$.\footnote{Besides, the theory
contains separately the Kaluza-Klein vector, but in the
reformulation (\ref{nonlin}) it appears only topologically.}
Therefore, the dual diffeomorphisms (\ref{diff}) give rise to an
additional symmetry,
$\delta_{\gamma}C_{3}{}^{3}=\partial_{3}\gamma^{3}$, which for
$\gamma^{3}=x^3\lambda$ implies the global shift symmetry
$\delta_{\lambda}\varphi=\lambda$ in the dimensionally reduced
theory. Thus, the $e$ transformations have been uplifted to $D=4$.
Unfortunately, the more interesting symmetries given by $f$ still
seem not to correspond to any invariance of the action
(\ref{nonlin}), in agreement with the essentially trivial
deformation of the gauge algebra analyzed above.

%%%%%%%%%%%%%%%%%%%%%%%%%%%%%%%%%%%%
\section{Comments and Outlook}\label{sec:parent}
%%%%%%%%%%%%%%%%%%%%%%%%%%%%%%%%%%%%

In this paper we have constructed a non-linear theory involving the
dual graviton. Instead of aiming at a non-abelian theory for the
dual graviton only --- which cannot exist in a local and covariant
fashion \cite{Bekaert:2002uh} ---, we derived a parent action, which
still contains the original metric. The latter guarantees invariance
under the full diffeomorphism group. However, this does not lead to
a doubling of degrees of freedom since there is no `kinetic'
Einstein-Hilbert term, while the metric enters through a topological
Chern-Simons-like term. Moreover, due to this topological term, the
theory can be shown to be classically equivalent to non-linear
Einstein-Hilbert gravity. It exhibits an enhanced gauge symmetry,
which contains not only the usual space-time symmetries, but also
`dual' diffeomorphisms and a local shift invariance. By virtue of
the shift gauge field, the non-linear duality relations between
metric and its dual are fully gauge covariant.

Thus, in total, we established the existence of a non-trivial theory
for the dual graviton, satisfying the requirements (i)--(iii) raised
in the introduction. One might wonder whether the necessity of
introducing a gauge field, which is a $(D-2)$-form with a Lorentz
index, has a natural interpretation within $E_{11}$. An inspection
of the relevant tables reveals that $E_{11}$ in the $SL(11)$
decomposition indeed has an $(D-2,1)$ Young tableau at level $7$
\cite{Nicolai:2003fw}, but that, at least at low levels, similar
objects seem not to appear for other decompositions or different
Kac-Moody algebras (as $A_{D-3}^{+++}$ in case of pure gravity)
\cite{Kleinschmidt:2003mf}. Thus, it is most likely that the shift
gauge fields have to be viewed as external quantities. This is not
an entirely unsuspected feature in that something similar happens
for the correspondence between gauged supergravity and $E_{11}$. In
fact, gauged supergravity requires the so-called embedding tensor,
which in turn is not predicted by $E_{11}$, but appears only through
its dual $(D-1)$-forms
\cite{Bergshoeff:2007qi,Bergshoeff:2007vb,Riccioni:2007au}. While
the latter, together with the $D$- or top-form potentials, encode
all constraints imposed by gauged supergravity, the embedding tensor
is nevertheless indispensable in order to construct an action
\cite{Bergshoeff:2007vb,Bergshoeff:2008qd}.

Unfortunately, the presented theory does not seem to fully uplift
the `hidden symmetries' of Kaluza-Klein reductions to the original
theory. This can be traced back to the fact that only the usual
diffeomorphisms are truly non-linear --- giving rise to the $SL(d)$
symmetry for reductions on $d$-tori ---, while the dual
diffeomorphisms are still abelian. Therefore, the symmetry
enhancement $SL(d)\rightarrow SL(d+1) (\rightarrow E_{d(d)})$ taking
place for reductions of (maximal super-)gravity can be elevated to
the higher-dimensional theory only for the positive-level `shift'
transformations. However, this is not different from the
correspondence between ordinary $p$-forms and Kac-Moody algebras.
(See the discussion in Sec.~3.1.) The results of this paper
therefore show that in this respect gravity is not special.

It would be interesting to extend this research into the following
directions. First of all, one might speculate that a true uplifting
of all hidden symmetries requires abandoning space-time covariance
as in \cite{Hohm:2005sc}. Moreover, even though we have seen that
generic matter couplings are compatible with the presented parent
action for dual gravity, it would be interesting to see whether for
special cases, like $3$- and $6$-form in $D=11$, an enhancement of
symmetries is possible such that the dual graviton starts
transforming under lower-level gauge transformations.

\section*{Acknowledgments} For useful comments and discussions we
would like to thank X. Bekaert, E. Bergshoeff, P. P. Cook, M. de
Roo, T. Nutma, H. Samtleben, P. Sundell and M. Vasiliev. The work of
N.B. is supported in part by the EU contracts MRTN-CT-2004-503369
and MRTN-CT-2004-512194 and by the NATO grant PST.CLG.978785. O.H.
is partially supported by the EU MRTN-CT-2004-005104 grant and by
the INTAS Project 1000008-7928. We are grateful to each other's
institutes for kind hospitality.

\end{document}